\documentclass[]{article}

\begin{document}
\begin{center}
\Large PROBLEMS WITH POPPER

\vspace{1.0cm}

\normalsize
{\it by Alan B. Whiting \\
University of Birmingham}
\end{center}

\vspace{1.0cm}

Professor Karl R. Popper was one of the most influential philosophers of
science in the twentieth century.  He is most widely known for his doctrine 
that scientific theories are not provable, but to be accepted as
scientific they must be falsifiable.  The most-cited statement of this is
from the {\em Postscript} to his {\em Logic of Scientific Discovery}$^1$:

\begin{quote}
\ldots we adopt, as our criterion of demarcation, the {\em criterion
of falsifiability, i.\ e.\ } of an (at least) unilateral or asymmetrical
or {\em one-sided} decidability.  According to this criterion, statements,
or systems of statements, convey information about the empirical world
only if they are capable of clashing with experience; or more precisely,
only if they can be {\em systematically tested}, that is to say, if
they can be subjected (in accordance with a `methodological decision')
to tests which {\em might} result in their refutation.  (pp.\ 313-4,
\S *i(2); his italics, as they will be henceforth)
\end{quote}

The book contains far more than this statement, however.
I set out some problems
with it below.

\section{Experimental uncertainty}

After asserting that every physical measurement is equivalent to noting
a pointer's position between two marks on a scale, which thus correspond to
an interval within which the measurement lies, Popper continues:

\begin{quote}
It is the custom of physicists to estimate the interval for every
measurement.  (Thus following Millikan they give, for example, the
elementary charge of the electron, measured in electrostatic units,
as $e=4.774 \ . \ 10^{-10}$, adding that the range of imprecision is
$\pm 0.005 \ . \ 10^{-10}$.)  But this raises a problem.  What can be
the purpose of replacing, as it were, one mark on a scale by 
{\em two}---to wit, the two bounds of the interval---when for each
of these two bounds there must again arise the same question:
what are the limits of accuracy for the bounds of the interval?

Giving the bounds of the interval is clearly useless unless these
two bounds in turn can be fixed with a degree of precision greatly
exceeding what we can hope to attain for the original measurement;
fixed, that is, within their own intervals of imprecision which should
thus be smaller, by several orders of magnitude, than the interval
they determine for the value of the original measurement.  In other
words, the bounds of the interval are not sharp bounds but are
really very small intervals, the bounds of which are in their turn
still much smaller intervals, and so on.  In this way we arrive at
the idea of what may be called the `unsharp bounds' or
{\em `condensation bounds'} of the interval. (p.\ 125, \S 37)
\end{quote}

There are two points I wish to make about this passage.  First, while
such a naive misconception of experimental uncertainty might be
understandable in someone who had never had contact with science at all,
it is
bizarre in a professor who is writing a book purporting to set out
very basic aspects of science.  Second, this misunderstanding generates a whole
conceptual structure (of `condensation bounds'), not only
taking up space in itself but developing further ideas (notably \S 68).

A significant fraction of the book is given over to criticism of the
quantum theory, in particular attempts to design thought-experiments to
disprove the Heisenberg Uncertainty Principle.  I will not go into them
in any detail here, since it would require a great deal of time and
attention; some criticism (too lenient, in my view) of his later 
writings on the subject has been published$^2$. 
But a taste may be
found on pp.\ 239-40, in which a monochromatic beam of particles has
been prepared (thus they are all of a single momentum); we are then, by
`focusing our attention' on those within an arbitrarily small volume, to
deterine position and momentum to any desired accuracy.  That Popper can
set this (explicitly `unphysical') procedure as a serious possibility
shows something of his incomprehension of the basics of experimental 
physics.

\section{Mathematics}

Most of the book is taken up with an analysis of the mathematical
theory of probability, including criticism of others' formulations
and a detailed presentation of Popper's own construction.  Before
treating probability as such, I want to look at a couple of examples
of Popper's use of mathematics.

In the first, Popper is concerned with the definition of a probability
apparently given by von Mises. 
In a sequence of events, the fraction
is formed of `successes' in which a particular thing occurs divided
by the total number of events.  If this fraction, called by Popper
the `relative frequency,' converges to a
definite number in the limit of an infinitely long sequence, that
number is the probability of success\footnote{Popper does not quote
von Mises explicity.  From the description it appears 
that the latter dealt with
`convergence in probability,' a phrase that Popper does not mention,
but which he would not have cared for.}.  The particular example used
is the fraction of ones in a sequence of ones and zeroes.

Popper wishes to do without the requirement of convergence.  That means,
he argues, he needs a concept that can be used in place of a limiting
frequency, applicable to all infinite sequences.

\begin{quote}
One frequency concept fulfilling these conditions is the concept of
a {\em point of accumulation of the sequence of relative frequencies.}
(A value $a$ is said to be a point of accumulation of a sequence if
after any given element there are elements deviating from $a$ by less
than a given amount, however small.)  That this concept is applicable
without restriction to all infinite reference sequences may be seen from
the fact that for every finite alternative {\em at least one} such
point of accumulation must exist for the sequence of relative 
frequencies which corresponds to it.  Since relative frequencies can
never be greater than 1 nor less than 0, a sequence of them must be
bounded by 1 and 0.  And as an infinite bounded sequence, it must 
(according to the famous theorem of Bolzano and Wierstrass) have
{\em at least one} point of accumulation. (p.\ 185, \S 64)
\end{quote}

The definition of accumulation point in the references I have at hand$^{3,4}$
is not the same, nor equivalent.  However, 
Popper's definition is probably contained within it,
so we will pass over that point.  His statement of the
Bolzano-Wierstrass theorem is accurate, except for the application
to a `finite alternative.'

The theorem only holds for infinite squences; in fact, a
finite sequence has no accumulation points$^3$.  In the paragraph
above, his mention of `finite sequences' suggests (at least) that he has not
grasped this requirement; this is proved on the next page (p.\ 186n4),
where he applies the concept of accumulation points to finite and
infinite sequences indiscriminantly.  Let me be explicit: Popper is
using a theorem in an area where it just isn't true.

The real problem with using accumulation points is that, while the
Bolzano-Weierstrass theorem assures us of at least one, there may be
many.  Popper realizes this (p.\ 185n2), and spends some time on the
rather trivial point that, in this case, they are not useful 
in defining probability.  He also recognises that to require
a unique accumulation point is equivalent to requiring convergence
(p.\ 186).  Then he requires uniqueness anyway, asserting that it isn't,
and in any case that he is free to choose such sections of any
sequence as have the behaviour he desires (p.\ 186n4).

I will pass over these last two problems with mathematical logic, because
in fact accumulation points are irrelevant to the task Popper attempts
in this section.  The importance of this episode is not so much that
Popper makes mistakes in the handling of accumulation points, which
could be considered a rather esoteric bit of analysis; nor even in his
failure to distinguish between finite and infinite, though that is
certainly a serious drawback for anyone trying to do mathematics.  It 
lies in the fact that he is not competent in this whole area of analysis,
deploying irrelevant machinery and doing that improperly.

Let us look at another section of mathematics, set theory.
Popper refers to Kolmogorov's development
of a theory of probability that explicitly uses sets.  But he does
not like it:

\begin{quote}
And yet, he [Kolmogorov] assumes that, in `$p(a,b)$'---I am using my own
symbols, not his [that is, the probability of a given b]---$a$ and $b$
are {\em sets}; thereby excluding, among others, the logical 
interpretation according to which $a$ and $b$ are statements (or
`propositions', if you like).  He says, rightly, `what the members of 
this set represent is of no importance'; but this remark is not
sufficient to establish the formal character of the theory at which
he aims; for in some interpretations, {\em $a$ and $b$ have no 
members}, nor anything that might correspond to members.  (p.\ 327,
\S *iv)
\end{quote}

Nowadays set theory is taught in elementary school; I am not sure what
its status was when Popper wrote.  But among the very first concepts
one comes across is $\emptyset$, the empty set, the set with no members.
Also among the first concepts is that a set may be made up of
{\em anything}, including statements, propositions, truth-values,
complex numbers, apples, oranges---or all at once.  I cannot escape the
conclusion that here Popper is criticising an approach knowing nothing
about it.  Indeed, in a later section (pp.\ 344-5, \S *iv) he
demonstrates the `superiority' of his `Boolean' approach over the
`set-theoretic' approach {\em by performing set operations}.

In presenting these two examples I am not asserting that Popper's mistakes and
misconceptions necessarily make all of his later work wrong.  That would
take a rather tedious effort of working through hundreds of pages of
sometimes convoluted logic.  I {\em am} asserting that he has attempted
to produce results with mathematics that he does not understand or,
worse, understands wrongly. 

\section{The probability of a hypothesis}

Popper takes issue with the idea that a theory, a hypothesis, may be
rendered more probable by a series of observations.
\begin{quote}
Let us now try to follow up the suggestion that the hypotheses themselves
are sequences of statements.  One way of interpreting it would be to take,
as elements of the sequence, the various singular statements which
can contradict, or agree with, the hypothesis.  The probability of the
hypothesis would then be determined by the truth-frequency of those 
among the statements which agree with it.  But this would give the
hypothesis a probability of 1/2 if, on the average, it is refuted
by every second singular statment of the sequence! (p.\ 257, \S 80)
\end{quote}

\noindent After considering a few modifications of this idea, he concludes,

\begin{quote}
This seems to me to exhaust the possibilities of basing the concept
of the probability of a hypothesis on that of the frequency of true
statements (or the frequency of false ones), and thereby on the
frequency theory of the probability of events. (p.\ 260, \S 80)
\end{quote}

There are two points to make about these statements immediately.
One is that Popper has set up an algorithm for the testing of a
hypothesis that is crude by any standard, and would not for a moment
be entertained by anyone actually attempting to test a hypothesis.
The second is that, in presenting this algorithm, he has set up
a `strawman,' any by refuting it has pretended to refute {\em all}
methods of testing a hypothesis (and giving it a greater or lesser
probability) by examining events\footnote{He may have derived this
formulation from an equation he attributes to Jeffries, p.\ 370, 
\S *vii, which he does not appear to understand.}.

Before returning to the question of the probability of a hypothesis
I want to bring out two other examples presented by Popper concerning
basic calculations in probability.  First, he wants to disprove the
`subjectivist theory of evidence,' what we would now call the Bayesian
approach (pp. 407-8, \S *ix).  Given a coin, what is the probability of 
heads?  Well, one-half.  Now given that the same coin gave 500,000
heads ($\pm 1350$) in one million trials, what is the probability?
One-half again.  Hence, under the subjectivist theory, after a million
coin flips {\em we have learned nothing}.

Popper believes this exercise disproves the idea of associating a
probability with a subjective belief in an outcome
(the initial guess of one-half), what we would now call a Bayesian
prior.  What he has actually done, by setting up the problem in a
way that no one with a background in even classical probability would
do, is prove that if an initial guess proves correct, we do not
change our minds\footnote{The more useful question is, before the
coin gets flipped, {\em how sure} are we of the initial guess of
probability one-half?  But there are many introductory texts that
consider the problem of the possibly biased coin, so I will not
go into more detail here.}.

Before we go on to the question of the absolute probability of a
theory, there is one section that I think illuminates Popper's
preconceptions in an interesting way.  On p.\ 390, \S *ix, he
deals with the throw of a die.  We take $p(x)$ as the probability of
throwing a six, $p(\bar{x})$ that of the negation (throwing anything
else).  Initially, with no information, we set the probabilities
as 1/6 and 5/6.  Then, we are given the information that that throw
is an even number.  The probabilities are now 1/3 and 2/3.  The
information has supported the hypothesis $x$ and weakened its
negation $\bar{x}$; but the fact that the probability of $x$ is still
smaller than $\bar{x}$ is considered by Popper `clearly self-contradictory,' and
thus proof that any calculation of the support or refutation of
a hypothesis (a theory) {\em cannot} be done within any
conventional probability theory.  Well, self-contradictory it is
not.  He appears to require that {\em any} support of a theory (a
hypothesis) make it more probable than its negation, a rather
absolute position.  He neither states this explicity nor attempts
to defend it (and defense it certainly requires), but it underlies
his work in this section.

We now come to Popper's calculation of the probability of a theory,
{\em any} theory, in the universe.  It is probably set out most
clearly on pp.\ 363-8, \S *vii.  The probability of a theory is
set equal to the product of the probabilities of all events
predicted by the theory, $p(a)= \prod p(a^n)$.  Since each component
$p(a^n)$ can never be greater than unity, may in fact be less, and
they will be infinite in number, the product will always go to zero.
Hence {\em the probability of any theory is exactly zero}.

I have already noted that this is not the way to work out
the probability of a hypothesis, given data.  It also assumes that
{\em all events in the universe are independent}.  Popper formally
recognises this, but asserts they must be, otherwise we could never
learn anything new (p.\ 368)\footnote{In passing, I note this logical
fallacy: Popper is asserting that if we do not know 
{\em everything}, we are required to know {\em nothing}.}.  
What it actually implies is total
chaos---no event would have any relation to any other, and from one 
moment to another, from one point to another, {\em anything} could 
happen.

Popper concludes something almost as depressing.  If $a$ is any theory
and $b$ is any information, {\em always}

\begin{eqnarray*}
p(a) & = & 0 \\
p(a|b) & = & 0 \\
\end{eqnarray*}

The consequences of this result will be traced in the next section.  
For the moment, let me emphasize that Popper's understanding of
elementary probability was inadequate and flawed, not
attaining the point of being able to test a possibly biased coin.

\section{Consequences}

Popper's conclusions to this point led him to construct a 
`calculus of relative probability' in which all of the
following formulae may be valid (p.\ 331, \S *iv):
\begin{eqnarray*}
p(a, b \bar{b}) &=& 1 \\
{\rm If}\ p(\bar{b}, b) \neq 0 && {\rm then}\ p(a,b) = 1 \\
{\rm If}\ p(a, \bar{a} b) \neq 0 && {\rm then}\ p(a, b) = 1
\end{eqnarray*}
These allow some very strange things to happen.  In the first, some
situation $a$ is certain to happen, given both $b$ and its negation.
That is, the given situation requires a single flip of a coin to be
{\rm both} heads and tails.  The second or third again allow a nonzero
probability to a situation and its negation: given that a die has
rolled a six, it is possible it rolled one through five.

In addition, one could also have simultaneously (for theories $a_1$ and $a_2$),
\begin{eqnarray*}
p(a_1, a_2) = 0 \\
p(a_2, a_1) = 1
\end{eqnarray*}
while at the same time $p(a_1) = p(a_2) = 0$ (p.\ 375, \S *vii).
Eventually he introduces a notion to express the fact that, while
the probability of every theory is zero, some zeroes are bigger
than others (pp.\ 375-6, \S *vii).

The kindest thing to say about a system that purports to do anything
with these statements is that it assigns meaning to something
essentially meaningless.  In fact one could say
Popper has succeeded in his goal of creating something that is not as `weak' as
conventional probability (p. 330, \S *iv).  By assigning a non-zero
probability to a situation in which a statement and its negation
are {\em both} true, his system can generate nonsense.
He does not appear to notice the inconsistency of this with his
claim to be implementing Boolean logic (p. 328, \S *iv).  In any case, it
is not something to be proud of.

\section{Summary}

In his most famous work, Karl Popper demonstrated an inadequate
and occasionally incorrect understanding of science, mathematics
and especially probability at an elementary level.  I have not here
looked at his philosophical ideas, but it would be surprising if
anything of much use could be built on such a base.

\vspace{1.0cm}
\begin{center}
{\it References}
\end{center}

\noindent (1) Karl R. Popper, {\em The Logic of Scientific Discovery}, 
including {\em Postscript: After Twenty Years}, (New York: Harper \& Row),
1968 (original edition published 1934, in English 1958) \\

\noindent (2) N. David Mermin, {\em Boojums All the Way Through}
Communicating Science in a Prosaic Age,
(Cambridge: Cambridge University Press), 1990, pp.\ 190-7 \\

\noindent (3) I. N. Bronshtein \& K. A. Semedyayev, {\em Handbook of
Mathematics}, English (third) edition, ed. K. A. Hirsch, (Heidelberg:
Springer-Verlag), 1998, p. 220 \\

\noindent (4) Norman B. Haaser \& Joseph A. Sullivan, {\em Real
Analysis}, (New York: Dover), 1971, pp. 82-4 \\
\end{document}